\documentclass{article}

\usepackage{color}
\def\be{\begin{equation}}
\def\ee{\end{equation}}
\def\lapl{\bigtriangledown}

\def\bear{\begin{eqnarray}}
\def\eear{\end{eqnarray}}
\begin{document}
\begin{center}
{\huge \bf Scaling in Complex Sytems: Analytical Theory of Charged Pores }\\
\vskip.1in
{A. Enriquez  and L. Blum }\\
\vskip.1in
{Department of Physics, University of Puerto Rico, P.O. Box 23343, Rio Piedras, PR , USA, 00931-3343}\\
\today \\
\vskip.1in

 {\bf \Large Abstract }\\ 
\vskip.1in
\end{center} 

 In this paper we find an analytical solution of the equilibrium ion distribution for a toroidal model of a ionic channel, using the  Perfect Screening Theorem (PST)\cite{blmg}.  The ions are charged hard spheres, and are treated using a scaling Mean Spherical Approximation (SMSA) \cite{vebl}.

Understanding ion channels is still  a very open problem, because of the many exquisite tuning details of real life channels. It is clear that the electric field plays a major role in the channel behavior, and for that reason there has been a lot of work on simple models that are able to provide workable theories. Recently a number of interesting papers \cite{ku0,ku1,ku2,jph1,jph2,jph3,beis1,beis2,beis3} have appeared that discuss models in which the effect of the geometry, excluded volume and non-linear behaviour is considered. 

We present here a 3D model of ionic channels  which consists of a charged,  deformable  torus with a circular or elliptical cross section, which can be flat or vertical (close to a cylinder).  Extensive comparisons to MC simulations were performed.\\

  The new solution, which is simple and accurate, opens new possibilities, such as studying flexible pores \cite{blube04} and  water turning phase transformations inside the pores, using an approach similar to that used on flat crystal surfaces \cite{tajk,c}

\section{Introduction}
We dedicate this contribution to Prof. Ben Widom, one of the leading figures in Statistical Mechanics.\\

  The study of the transport process in membrane ion channels is complicated by the presence of the protein walls, the interaction with ions, water molecules and the electric field  profile, which determine many of the salient properties of ion channels \cite{ku2}. Computing the electric potential profile everywhere in a real channel is difficult, if not impossible, because of the complexity of the system, and for that reason simplified models have been used: Kuyucak et al \cite{ku1,ku2},  have studied circular toroidal channels using the linear Poisson-Boltzmann equation, in which the ions are treated as point charges.  
Excluded volume effects have been included  \cite{jph3} to explain ion selectivity.
Furthermore nonlinear effects, which are important,  are included in the 1D, non-linear Poisson Boltzmann PNP models of Eisenberg et al.\cite{beis1,beis2,beis3}. Excluded volume effects come into play when molecular solvents are used\cite{jph1,jph2}. Recently  the effects on porin arising from the  rotation of water molecules were discussed\cite{tajk}.\\

We propose here a  SMSA  solution \cite{vebl} of a ion channel model  which consists of a toroidal ring
 with either circular or elliptical (prolate or oblate) cross section. The solution is given in terms of a few MSA scaling parameters provide an accurate description of the charge disribution and  can be obtained from a variational theory. The major feature of our solution is that it satisfies the Perfect Screening Theorems (PST)\cite{blmg}. This is not only a physical requirement, but also a technical advantage, because it can used to include discrete molecular solvents such as hard dipoles \cite{id02}, and  water \cite{yukag}. This has  been used in the theory of a phase transition that involves the turning of water by an electric field\cite{c}.

The non-linear Poisson-Boltzmann case has been discussed elsewhere \cite{adv80,beblu00} and as a matter of fact, is implicit in our present work.

 To study the dynamics of ions in a channel, one needs to compute the forces acting on each of the ions , including mobile, induced and fixed charges  and the applied electrical field. This could be coupled with Brownian dynamics simulations or other coarse grained simulations. Because this computation has to be repeated at every step, the existence of analytical solutions in a relevant geometry is imperative for simulations at realistic time scales. 

\subsection{The perfect screening sum rule (PST)}

 One 
remarkable property of mixtures of classical charged particles is that 
because of the very long range of the electrostatic forces, they must 
create a neutralizing atmosphere of counterions, which shields 
{\em perfectly} any charge or fixed charge distribution.
Otherwise the partition function, and therefore all the thermodynamic functions, will be 
divergent \cite{blmg}. This sum rule is intuitive and widely accepted for spherical systems. But it is also true for non-spherical systems. It has been explicitly verified  for non spherical systems in simulations and also using exact results for the Jancovici model  \cite{robl,sabl}. \\

For spherical ions this means that the internal energy $E$ of the ions is always the sum of the energies 
of {\it spherical} capacitors. For {\it any} approximation  
the {\it exact} form of the energy is
\be
 \Delta E=-\frac{e^2}{\varepsilon} 
\sum_i\rho_{i}z_i  \frac{ z_i}{1/\Gamma_i+\sigma_i},
\label{eq1}
\ee 
 $\beta = 1/kT$ is the usual Boltzmann thermal factor, $\varepsilon$ 
 is the dielectric constant, $e$ is the elementary charge, and ions $i$ 
have charge, diameter and density $z_i$, $\sigma_i$, $\rho_i$, respectively. $\Gamma_i$ is the shielding length for ion i. \\

An equivalent electrostatic model of eq.(\ref{eq1}) is the ion immersed in a conducting media, which means that the energy will depend only on the charge of the ion and not on its internal distribution: The energy will be the same if the charge is concentrated at the center of the sphere or uniformly distributed over the surface of the ion \cite{yrbl1,yrbl2,yrbl3}: Clearly the potential at the surface has to be constant and the proper boundary conditions for the electrostatics are the Dirichlet boundary conditions. The same considerations apply for an {\it arbitrary}  fixed charge distribution. In the case of a solid toroid of circular section immersed in a conducting media the potential outside of the torus will be the same if we took a circular wire at the center  of the torus or a charged ring with a constant surface potential. We remark that this is true strictly for Dirichlet boundary conditions. The more general case is more complicated.\\

For non spherical systems the perfect screening theorem requires that all multipoles 
of the fixed charge distributions be compensated by the mobile charge distribution. As we will see below this implies a very substantial simplification of the solution of linear PB equation since every multipole of the countercharges distribution cancels the fixed charges multipoles and all cross terms are zero.

\section {The charged torus}

Analytical solutions of closures of the Orstein-Zernike (OZ) or Wertheim-Ornstein-Zernike (WOZ) equations are only possible in odd parity spaces. The torus is an  even parity,  2-dimensional object, and for that reason there is no direct analytical solution possible of the MSA or LPBE. The expansion in spherical harmonics on the other hand is always possible  and for our model is convergent within a reasonable (even small!) number of spherical harmonics
\cite{cohltol}. The beauty of the PST is that it de-couples all the multipole terms, and for that reason we  are able to solve the LPBE to all orders in closed form. 

Consider first a circular section $b=b_{real}+\sigma/2$ torus of radius $a$. The diameter of the ions is $\sigma$. The electrostatic equivalent system is a ring wire of radius $d=a$ and the torus  immersed in a conducting media.  
Poisson's equation for the potential in a charged system is \cite{1}
\be
\nabla ^2 \phi({\bf r})=\frac {4\pi}{\epsilon}  q({\bf r}),
\label{eq1}
\ee
Here the charge density $q(r)$ at r is the sum of the fixed ring   and the mobile ion charges
\be
q({\bf r})=q_{ring}({\bf r})+\sum_i q_i({\bf r})
\label{eq3}
\ee
\noindent where $\phi({\bf r})$ is the potential  at ${\bf r}\equiv R,z$ .\\ 

 The formal solution of this equation is
\be
\phi_({\bf r})=\frac{1}{\epsilon} \int d^3{\bf r}' \frac{q({\bf 
r}')}{|{\bf r}-{\bf r'}|}=\frac{1}{\epsilon} \int d^3{\bf r}' \frac{q_{ring}({\bf 
r}')}{|{\bf r}-{\bf r'}|}+\frac{1}{\epsilon}\sum_i \int d^3{\bf r}' \frac{q_i({\bf 
r}')}{|{\bf r}-{\bf r'}|}
\label{eq4}  
\ee

 The potential generated by a charge Q on the ring  of radius  $d$ is given by \cite{lebed,cohltol}
\bear
\phi_{ring}(R,z)=\frac{2 Q}{\epsilon[(R-d)^2+z^2]} K(-m);\qquad m\equiv\frac{4 R d}{(R-d)^2+z^2}
\label{eq5}
\eear
where $K(m)$ is the elliptic function
\be
K(m)=\int_0^{\frac{\pi}{2}} \frac {d\phi}{\sqrt{1-m^2 \sin^2 {\phi}}}
\ee
which satisfies the  homogeneous Poisson equation

\be
\lapl^2 \phi=0.
\ee
However, the inhomogeneous Poisson equation (\ref{eq1})
 has no closed form analytical solution.  We need to  expand our problem in a suitable basis. We use spherical harmonics expansion  because its good analytical behavior (it has been extensively used in astrophysics \cite{cohltol}), but more important, because of the PST \cite{blmg} the terms are decoupled to each order in the expansion. For the ring source potential there are three regions which correspond to  outside and inside  a sphere of radius $a$ of the ring.

\bear
\phi_0(r)&=&
\phi^{ext}\theta_{Heav}(r-a-b)+\phi^{ring}\theta_{Heav}(a+b-r)\theta_{Heav}(r-a+b)\nonumber \\&+&\phi^{int}\theta_{Heav}(a-b-r)
\nonumber\\
\label{eq8}
\eear
The parameter b is the effective diameter of the circular torus. We assume that the ions are of diameter $\sigma$ and therefore the real radius of the toroid is
\[
b_{real}=b-\frac{\sigma}{2}
\]
When the width of the ring is $b=0$ we get

\be
\phi^{ext}(r)=\sum_{\ell=0}^{\infty} P_{\ell }(\cos  \theta 
)r^{-(\ell+1)} M_{\ell}^{ext}
\label{eq9}
\ee
with
\be
M_{\ell}^{ext}=\frac{Q}{\epsilon} P_{\ell}(0)a^{\ell}.
\ee

and
\be
\phi^{int}(r)=\sum_{\ell=0}^{\infty}r^{\ell} P_{\ell }(\cos  \theta 
) M_{\ell}^{int}
\ee
with
\be
M_{\ell}^{int}=\frac{Q}{\epsilon} P_{\ell}(0)a^{-(\ell+1)}.
\ee
The fixed multipole moments $M_\ell$ are the same for a ring of charge $Q$ or a solid toroid with uniform potential on its surface ( Dirichlet boundary conditions). This will be true for a pore immersed in a conducting electrolyte. We will take advantage of this fact computing the moments for the ring of zero with, and then using them in the calculation of the potential for a toroid of finite width.\\
 
The calculation of the multipole moments $M_\ell$ for the general case of elliptical toroids is left for a future publication.\\

$\,\,$\\\\\\\\\\\\\\\\\\\\\

\includegraphics{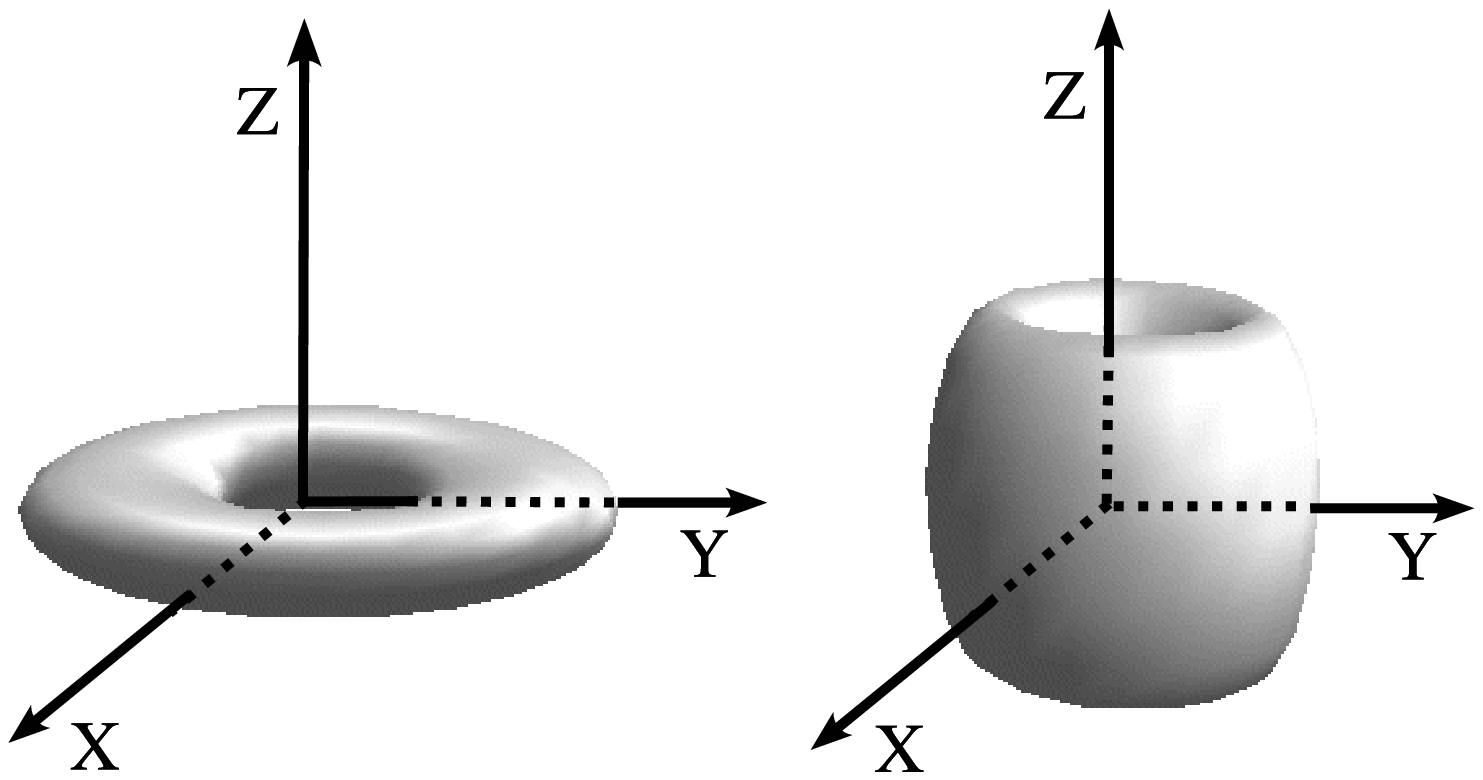}

\hspace{1cm} {\bf Fig.1a.} Charged rings  with elliptical cross sections.\\ 

$\,\,$\\\\\
$\,\,$\\\\\\\\\\\\\\\\\\\\\

\includegraphics{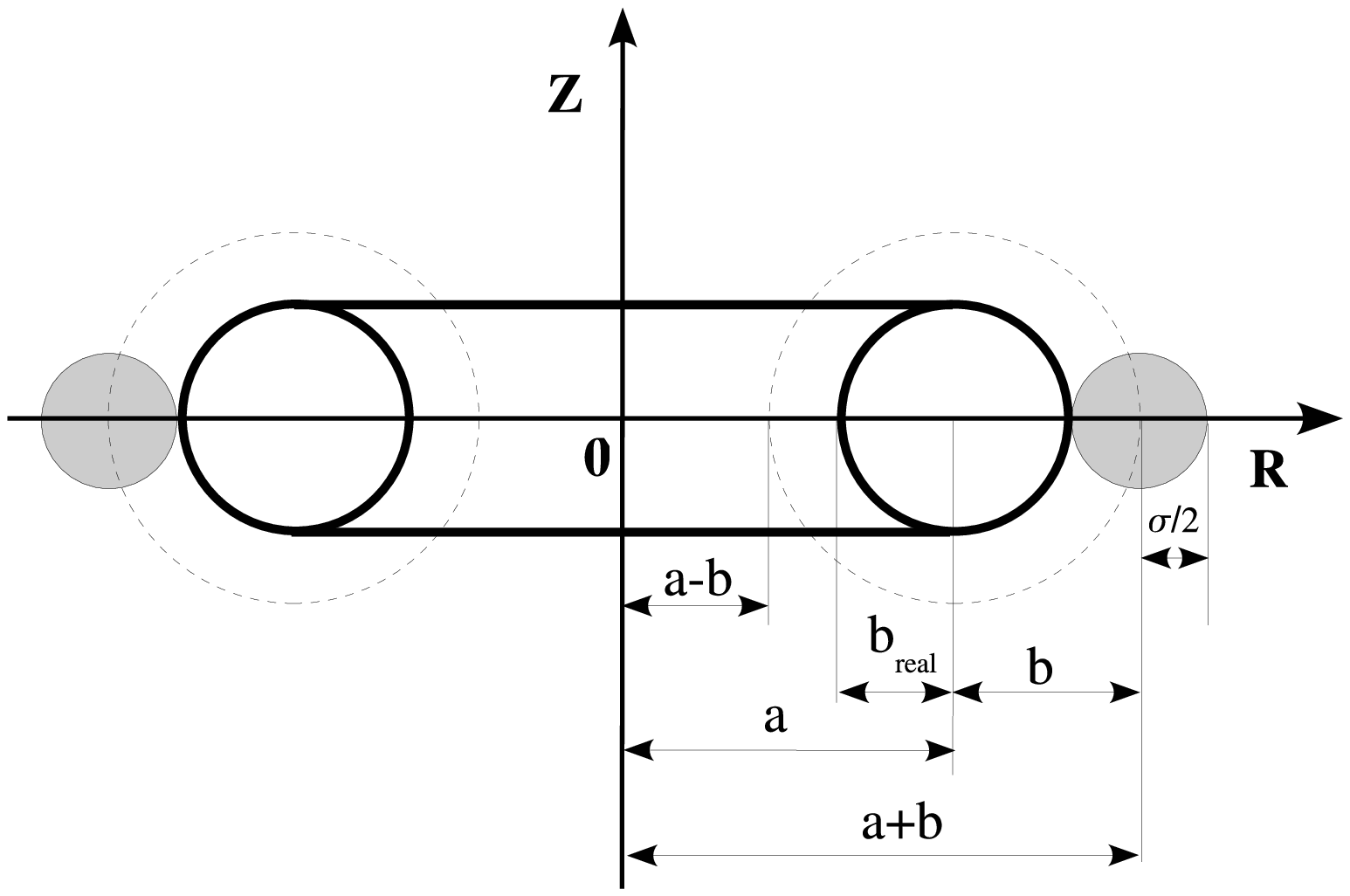}

\hspace{1cm} {\bf Fig.1b.} \\

\section{The Solution of the Linear Poisson Boltzmann Equation (LPBE) and the Scaling MSA}

But the LPBE is accurate only for very dilute solutions because the ions are point charges. The MSA is the LPBE, but with the mathematically correct treatment of the excluded volume effects.
  The MSA which is derived  from the Mean Spherical Model \cite{percus-yevick,lebowitz-percus,waisman-lebowitz,ewaisman}, provides a  coherent and simple description of the properties of a large class of systems   in terms of a very small set of scaling parameters $\Gamma_\alpha$ This includes ionic solutions, water and polyelectrolyte \cite{lb1,bh,b80,bh77,gin1,bvh92,blmhe3,blmub,bhol03}. The functional form of the thermodynamic parameters in the different modified MSA theories is the same as in the LPBE, however  the excluded volume is treated exactly in the MSA  which satisfy the Onsager high density bounds \cite {yr1,yrbl2,yr3}, and is  asymptotically exact at large concentrations.
 One can show that simple transformations lead to the proper high density behavior. The Debye screening length $\kappa$ in the DH theory becomes the MSA screening length $\Gamma$

\be
\kappa\equiv\sqrt{\frac{4 \pi \beta e^2}
{\varepsilon}\sum_{j=1}^m \rho_j  z_j^2}\quad{\bf\Longrightarrow}\quad\Gamma \equiv\frac{1}{2 \sigma}\sqrt{(1+2\kappa \sigma)}-1
\label{eq13}
\ee

It can be shown that the proper high density behavior in the MSA stems  from the fact that the entropy is of the form
\be
\Delta S^{(MSA)}= -k_B\frac{ \Gamma^3}{3 \pi}
\label{ds}
\ee
where $k_B$ is the Boltzmann constant.
For nonspherical systems this generalizes to\cite{id02}
\be
\Delta S^{(MSA)}= -k_B\sum_{\chi=-\ell}^\ell\frac{ \Gamma_\chi^3}{3 \pi}
\label{eq15}
\ee
where $\chi$ is the index of the irreducible representation \cite{lb73} and $\ell$ is the order of the spherical harmonic in eq.(\ref{eq9}).
 This immediately suggests \cite{yr3,vebl}  that $\Gamma_\chi$ can be  determined 
by the variational expression
\be
\frac{\partial[\beta \Delta A_\chi(\Gamma_\chi)]}{\partial \Gamma_\chi}=\frac{\partial[\beta \Delta E_\chi(\Gamma_\chi)+\Gamma_\chi^3/(3 \pi)]}{\partial \Gamma_\chi}=0
\label{eq16}
\ee
 For the simple restricted case of an equal size ionic mixture we get equation (\ref{eq13}).  For more complex systems, like the general polyelectrolyte this equation is a new relation to be found. For flexible polyelectrolytes it has been derived from the binding MSA (BIMSA) \cite{yr3,bernard-blum,blukab}.\\

In this work we use the LPBE in conjunction with the perfect screening theorem (PST)\cite{blmg}, to derive the functional form of the solution to our problem. This functional form provides an astonishingly simple and  good representation of our simulation data. From  eq.(\ref{eq16}) we get a very good first approximation to the charge distribution obtained from our extensive simulations. 
\\

\subsection{Solution of the LPBE}
Consider eq.(\ref{eq4}) : In the linear Poisson Boltzmann approximation we write
as a convolution:
\be
\phi(r)=\phi_0 - \frac{\kappa^2}{4\pi}\left (\frac{1}{r}\right)*\phi(r)
\ee
Taking the  Fourier transform of  both sides
\be
\widetilde{\phi}(k)=\widetilde{\phi}_0-\frac{\kappa^2}{4\pi}\widetilde{\left(\frac{1}{r}\right)}\widetilde{\phi}(k)
\ee
where
\be
\widetilde{\left(\frac{1}{r}\right)}=\int 
d^3r\frac{e^{i\vec{k}.\vec{r}}}{r}=\frac{4\pi}{k^2}
\ee
Then,
\be
\widetilde{\phi}(k)=\frac{k^2}{k^2+\kappa^2}\widetilde{\phi}_0
\label{eq20}
\ee
where
\be
\widetilde{\phi}_0=\widetilde{\phi}_0^{int} +  
\widetilde{\phi}_0^{ext}+  
\widetilde{\phi}_0^{ring}
\label{eq21}
\ee

\be
\widetilde{\phi}(k)=\frac{k^2}{k^2+\kappa^2} 
\widetilde{\phi}_0^{int} + \frac{k^2}{k^2+\kappa^2} 
\widetilde{\phi}_0^{ext}+ \frac{k^2}{k^2+\kappa^2} 
\widetilde{\phi}_0^{ring}
\ee
Using Rayleigh's formula 
\[
e^{i k r \cos\theta}=\sum_\ell (2 \ell+1)i^\ell P_{\ell}(cos\theta)j_{\ell}(k r)
\]
we get
\be
\widetilde{\phi}_0^{int}=4\pi \sum_{\ell=0}^{\infty} 
M_{\ell}^{int} P_{\ell}(cos\theta) i^{\ell} \int_0^a j_{\ell}(k r)r^{\ell+2} 
dr,
\ee
and

\be
\widetilde{\phi}_0^{int}=4\pi\sum_{\ell=0}^{\infty} 
M_{\ell}^{int}\ P_{\ell}(cos\theta)\ i^{\ell}\ \frac{a^{\ell+2}}{k}\ 
j_{\ell+1}(k a).
\ee

\noindent  We find $\widetilde{\phi}_0^{ext}$ from 
equation (13):

\be
\widetilde{\phi}_0^{ext}(k)=4\pi 
\sum_{\ell=0}^{\infty}M_{\ell}^{ext}\ P_{\ell}(cos\theta)\ i^{\ell}\ \frac{a^{1-\ell}}{k}\ 
j_{\ell+1}(k a).
\ee
and $\widetilde{\phi}_0^{ring}$ is a linear combination of $\widetilde{\phi}_0^{ext}$ and $\widetilde{\phi}_0^{int}$ with coefficients to be determined by the boundary conditions.

The inverse  Fourier transform of $\phi^{ext}(r)$   is
\be
\phi^{ext}(r)=4\pi\sum_\ell P_{\ell}(cos\theta) 
R_{\ell}^{ext}(r),
\ee
where, after some calculations \cite{6}
\bear
 R_{\ell}^{int}(r)&=& \sum_{\ell=0}^{\infty} M_{\ell}^{int} a^{\ell+2} 
i^{\ell}\frac{1}{(2\pi)^3}\int_0^{\infty}d^3k 
\frac{k}{k^2+\kappa^2}j_{\ell+1}(k a)e^{-i\vec k.\vec r}\nonumber \\
&=&M_{\ell}^{int}\frac{2}{\pi} a^{\ell+2}\int_0^{\infty}dk 
\frac{k^3}{k^2+\kappa^2}j_{\ell}(k r)j_{\ell+1}(k a)
\nonumber \\
&=&M_{\ell}^{int}\kappa^2\ a^{\ell+2}\ i_{\ell}(\kappa r)\ k_{\ell+1}(\kappa a)\nonumber \\
\label{eq29}
\eear
Similarly
\bear
 R_{\ell}^{ext}(r)&=& \sum_{\ell=0}^{\infty} M_{\ell}^{ext} a^{1-\ell} 
i^{\ell}\frac{1}{(2\pi)^3}\int_0^{\infty}d^3k 
\frac{k}{k^2+\kappa^2}j_{\ell-1}(k a)e^{-i\vec k.\vec r}
\nonumber \\
&=&M_{\ell}^{ext}\kappa^2\ a^{1-\ell}\ k_{\ell}(\kappa r)\ i_{\ell+1}(\kappa a)\nonumber \\
\label{eq30}
\eear

\noindent Finally we find 
the LPBE potential
\bear
\phi^{ext}(r)&=&\sum_{\ell}P_{\ell}(cos\theta)[  
B_{\ell}\ k_{\ell}(\kappa r)]\nonumber \\
\phi^{int}(r)&=&\sum_{\ell}P_{\ell}(cos\theta)[  
A_{\ell}\ i_{\ell}(\kappa r)]\nonumber \\
\phi^{ring}(r)&=&\sum_{\ell}P_{\ell}(cos\theta)\left\{  
C_{\ell} i_{\ell}(\kappa r)+D_{\ell} k_{\ell}(\kappa r)\right\}\nonumber \\
\label{eq29}
\eear
where
\be
A_{\ell}= M_{\ell}^{int}\ \kappa^2\ a^{\ell+2}\ k_{\ell+1}(\kappa a),
\ee
\be
B_{\ell}= M_{\ell}^{ext}\ \kappa^2\ a^{1-\ell}\ i_{\ell+1}(\kappa a)
\ee
and $C_{\ell}$ and  $D_{\ell}$ are found from the boundary conditions on the spheres of radius $r=a\pm b$.
The potential at the interface is  constant, so that in the simplest case they are obtained by matching the normal derivatives at the toroid's surface. This means that the matching at the toroid's surface must  include a geometrical factor
\bear
b(z)&\equiv &\sqrt{b^2-z^2} ; \qquad b>z
\nonumber\\
\cos{\theta}&=&\frac{z}{\sqrt{a^2+b^2\pm 2 a b(z)}}
\label{eq32}
\eear
This factor only matters when $z\approx b$.\\

From the simulations we know these coefficients depend also on nonlinear effects. But what is clear is that they provide a good representation of the results of our current simulations. \\

The spherical bessel functions $i_\ell(\kappa r)$ and  $k_\ell(\kappa r)$ \cite{abram} satisfy the proper boundary conditions :

\be
\phi_{int}(r,\theta)=\sum_{\ell}A_{\ell}\ P_{\ell}(cos\theta)\ 
i_{\ell}(\kappa r),\qquad(0<r<a-b)
\ee
The exterior potential  is
\be
\phi_{ext}(r,\theta)=\sum_{\ell}B_{\ell}\ P_{\ell}(cos\theta)\ 
k_{\ell}(\kappa r),\qquad(r>a+b)
\ee
As was discussed elsewhere \cite{vebl,blube04,yr3} it has been shown that in the many cases where 
an analytical solution of the MSA, for complex systems is available, the solution can be obtained from a simple variational principle. The actual solution of the MSA  is more complex for higher values of $\ell$. But in our simulations the largest term is the one with $\ell=0$ so that eq.(\ref{eq13}) is a quite decent approximation.  
Then our  expression is
\bear
\phi_0(R,z)=
\phi^{ext}\theta_{Heav}(r-a-b(z))+\phi^{int}\theta_{Heav}(a-b(z)-r)\nonumber \\+\phi^{ring}\theta_{Heav}(a+b(z)-r)\theta_{Heav}(r-a+b(z))
\nonumber\\
\label{eq34}
\eear  
with
\bear
\phi^{ext}(r)&=&\sum_{\ell}P_{\ell}(cos\theta)[  
B_{\ell}\ k_{\ell}( \Gamma r)]\nonumber \\
\phi^{int}(r)&=&\sum_{\ell}P_{\ell}(cos\theta)[  
A_{\ell}\ i_{\ell}(\Gamma r)]\nonumber \\
\phi^{ring}(r)&=&\sum_{\ell}P_{\ell}(cos\theta)\left\{  
C_{\ell} i_{\ell}(\Gamma r)+D_{\ell} k_{\ell}(\Gamma r)\right\}\nonumber \
\eear
where
\be
A_{\ell}=4\pi M_{\ell}^{int}\Gamma ^2\ a^{\ell+2}\ k_{\ell+1}(\Gamma a),
\ee
\be
B_{\ell}=4\pi M_{\ell}^{ext}\Gamma ^2\ a^{1-\ell}\ i_{\ell+1}(\Gamma a)
\ee

\section{Computer Simulations}

The geometry of the system is displayed in Figure 1. We use a cylindrical ring of radius $b=1/2$ The size of the ions is also $\sigma=1$.  We have chosen this model because it is closest to actual channel geometry 
\cite{ku0,ku1}. As is shown in figure 1 we can stretch or flatten the torus so that it can be a narrow channel or just a pore in a flat membrane.  In our solution this means that we merely have to increase the number of multipoles in the expansion eq.(\ref{eq9}). As was shown by Tohline et al. the expansion of toroidal systems using spherical harmonics can be carried out up to quite large values of $\ell\simeq 1000$ \cite {cohltol}. Simulations discussed in this paper are only for cylindrical toroids. We simulate a cylindrical section torus for which only a few terms are needed (typically 3 or 4). We use standard Monte Carlo simulation techniques \cite{all,fren,sad}.  Our entire system is confined to a large  sphere, and our charged ring is lying flat in the x-y plane, so that $z=0$ is the plane of the ring. Typically the radius of the sphere is 20-26 times the ionic diameter, although runs have been performed with larger spheres to assure that the influence of the walls is  small. The number of ions varied between 200 and 512. The sampling bins were cylindrical sections, which makes the data near the axis of the torus particularly noisy. For that reason we have used extremely long runs ( $10^8$ steps  in some cases) and even so we had to discard the noisy central region. Independence of the size of the spherical container was verified.\\

In  the simulations the ring had a radius of 3, and since the ions had a radius of 1, the effective exclusion region goes from 2 to 4 units. 
The results of figures 2 and 3 for the charge distributions in the x-y plane
 are well represented by eq.(\ref{eq34}), using the first two terms in the expansion.
Figures 4 and 5 represent the same charge distributions for different values of z. The curve for z=1 marks the transition region where the ring ends. The curve is a perfect parabola given by eq. (32) and spans the region $a\pm b(z)$. For $z>b=1$ the profile is simply a superposition of the interior and exterior regions.
The overall agreement of these figures with the theory is comparable to those of figures 2 and 3.

Figures 6 and 7, which correspond to $z=0$, and charges $Q=4,12,20$, are also in excellent agreement with the theory of eq. (34), in the sense that the functional form is always the same. However the parameters $A_\ell,B_\ell,C_\ell,D_\ell$ , have a nonlinear dependence on the charge Q. We hope to discuss this point in the near future.\\
 
Our simulations are designed to satisfy the perfect screening theorem,\cite{blmg}, and therefore we will not have large multipoles ($\ell\ge 2$) induced by the periodic boundary conditions.
\newpage
\includegraphics{sian200r3ch20an.eps}

$\,\,$\\\\\\\\\\\\\\\\\\\\\\\\\\\\\\\\

\hspace{0cm} {\bf Fig.2.} Comparison of theory and computer simulation of  the anions density for a ring of internal diameter 3, charge 20. The dotted line is the theory.\\

\includegraphics{sian200r3ch20ca.eps}

$\,\,$\\\\\\\\\\\\\\\\\\\\\\\\\\\\\\\\\\\\\\\\\\\\\\\

\hspace{0cm} {\bf Fig.3.} Comparison of theory and computer simulation of  the cations density for a ring of internal diameter 3, charge 20. The dotted line is the theory.
 
\newpage

\includegraphics{figura4.eps}

$\,\,$\\\\\\\\\\\\\\\\\\\\\\\\\\\\\\\
\hspace{0cm} {\bf Fig.4.} $g_{cation}(R,z)$ for  the same parameters as in figure 3. The graph clearly shows the three regions of eq.(\ref{eq21}) when $z<b$ and the two regions for  $z\ge b$

\includegraphics{figura5.eps}

$\,\,$\\\\\\\\\\\\\\\\\\\\\\\\\\\\\\\\\\\\\\\\\\\\\\\\\\

\hspace{0cm} {\bf Fig.5.} $g_{anion}(R,z)$ for  the same parameters as in figure 4. The graph also shows  three regions of eq.(\ref{eq21}) when $z<b$ and the two regions for  $z\ge b$

\newpage

\includegraphics{figura6.eps}

$\,\,$\\\\\\\\\\\\\\\\\\\\\\\\\\\\\\\

\hspace{0cm} {\bf Fig.6.} Charge dependence of  
       $g_{anion}(R,z)$ for  z=0.

$\,\,$\\\\\\\\\\\

\includegraphics{figura7.eps}

$\,\,$\\\\\\\\\\\\\\\\\\\\\\\\\\\\\\\\\\\\\\\
\hspace{0cm} {\bf Fig.7.} Charge dependence of  
       $g_{cation}(R,z)$ for  z=0.

\newpage

\section {Discussion of Results}
 
This research is part of an ongoing effort to describe complex systems by means of a small set of scaling parameters.
The basic idea is to use a singlet density description combined with exact relations, such as the Onsager asymptotic limits, or the Wertheim association limits or, as in the present case, the perfect screening theorem.
For example we have developed a  theory of flexible polyelectrolyte of arbitrary length that satisfies an exact relation for infinite length\cite{blube04,beblu00}, and for that reason agrees extremely well with computer simulations. The same theory has been used to describe a completely flexible ring (Bernard and Blum, unpublished): In this case the 'channel' is a chain of spherical beads of arbitrary charge and diameter. The toroidal channel is a model that in a certain sense represents the other extreme situation since it is rigid and structure-less. The 'true' model lies somewhere in between.

Another situation where our solution could be used is to describe the case when the water molecule turns, as was discussed by Tajkhorshid et al. \cite{tajk}. In fact we have used our methods to describe a phase transition which is actually due to the turning of water in an external electric field \cite{c}. 

Also dielectric effects  can be included, adding another scaling parameter that has the meaning of a polarizability. This was recently published \cite{id02}. Actually equation (16) is derived in that paper. And finally, external fields are easy to include and the effect of membranes can be mimicked by a suitable deformation of the torus shown in figure 1.

We have presented here a theory of the equilibrium distribution of ions in a charged channel that is surprisingly simple. The fact that so few parameters can describe the distribution of charges inside and outside the channel is very appealing. A feature of our approach is that it can be easily extended to include polarization effects, solvent and discrete structures. Because of the availability of an analytical solution of the octupolar model of water\cite{c} it is another definite possibility.\\

\section{Acknowledements}
Support from  DOE through grant DE-FG02-03ER 15422 is  graciously acknowledged. We acknowledge the invaluable help of Profs. A.Z. Panagiotopoulos, P. Moore and Bernd Ensing in the developement of our MC code. Parts of this research was done at Princeton University, and we thank Prof. Pablo Debendetti for his hospitality there. We also enjoyed the hospitality of Prof. M.L. Klein at the LRSM of the University of Pennsylvania, and L.B and A.E. the support through  Grants NSF-DMR-9872689 and DMR-0353730.\\

Very useful suggestions were provided by Prof. Angel Garcia (formerly at LANL, currently at RPI) ,
\\

\end{document}